\documentclass[doublecol]{epl2} 
\usepackage{amsmath}
\usepackage{hyperref}
\title{Microwave-Dressing of Rydberg States in a Trapped Calcium Ion}
\shorttitle{Microwave-Dressing of Rydberg States in a Trapped Calcium Ion} 

\author{H. Bao\inst{1} \and A. Schulze-Makuch\inst{1} \and F. Schmidt-Kaler\inst{1,2}}
\shortauthor{H. Bao, \etal}

\institute{                    
  \inst{1} QUANTUM, Institut f\"{u}r Physik, Universit\"{a}t Mainz, D-55128 Mainz, Germany\\
  \inst{2} Helmholtz-Institut Mainz, D-55128 Mainz, Germany
}
\pacs{32.80.Ee}{Rydberg states}
\pacs{33.40.+f}{Multiple resonances}
\pacs{37.10.Ty}{Ion trapping}

\abstract{We are using optical- and microwave-fields to excite Rydberg states in trapped cold $^{40}Ca^+$ ions. We employ a single ion and observe spectroscopically in the manifold of a principal quantum number n=49 the dressing of Rydberg states of angular momentum states S and P. We compare our experimental spectra with a multi-level calculation of dressed states and find good agreement. The results are important for controlling the interaction of single ions in Rydberg states with electric fields of the ion trap, and for tailoring the interactions in an ion crystal in Rydberg states.}

\begin{document}
\maketitle

\section{Introduction and Motivation}
Rydberg atoms feature unique properties, states where the valence electron is excited to a high principal quantum number n. The atomic wavepacket size increases with n$^2$, and the electric dipole moment correspondingly scales with n$^4$. Fritz London worked out the quantum mechanical theory of dispersion forces between instantaneously induced dipoles, fluctuating in time, but still leading to a van der Waals potential. For trapped Rydberg ions, this van der Waals interaction remains rather small: For a pair of Sr$^+$ ions at a distance of 4$\mu$m and with n=50, we calculate an energy J$\simeq$~8.5~kHz. However, if the interaction is not between fluctuating dipoles, but rather a Debye-force between permanently induced dipoles, the interaction strength increases by almost three orders of magnitude to J$\simeq$~3.1~MHz. For neutral Rydberg atoms, such dipole dipole interaction can be tuned on by employing the Stark effect, as studied early on by Daniel Kleppner~\cite{Klepp1995} and Thomas F. Gallagher\cite{GALLAGHER2008161}. In neutral atoms a static electric field is used for mixing manifolds of Rydberg states, such that a giant dipole-dipole interaction is induced. For atomic ions confined in a Paul trap, the electric trapping potential in x, y, and z direction is harmonic with the ion confined at the center of the potential. Applying a static additional electric field will only modify the center of the harmonic electric potential,  such that the ion wavepacket is shifted to a new equilibrium position at zero electic field strength. Thus, a static electric field is not suited to induce a dipole moment in the Rydberg manifold of a trapped ion. 

Previous works\cite{Zha2020} with a pair of trapped Sr$^+$ ions has demonstrated that a microwave electric field can be used for dressing Rydberg S- and P-manifolds, to turn on a dipole-dipole interaction. This interaction has been used to achieve a high speed entangling gate, see also Ref. ~\cite{Li2013}. The observed infidelity of 23 $\%$ of this two-ion gate operation~\cite{Zha2020}, however, was dominated by power instabilities when applying the microwave field. Indeed, the microwave-dressed Rydberg state energy levels depend on the microwave Rabi frequency, such that an excellent power stability is important for a high fidelity gate operation. In this letter, we employ a single Ca$^+$ ion and apply laser excitation to a Rydberg state, while dressing the states by an injected microwave electric field. We observe the Autler-Townes splitting, and compare our experimental results with a calculation. This comparison is indispensable for understanding how the microwave field with a wavelength of about 3.3~mm interacts with the ion sitting at the center of a trap structure, that features metallic electrodes at a much smaller distance of 0.75~mm. We conjecture that the injected TEM(0,0) mode MW beam becomes strongly affected by near-field effects. This reasoning fits to our observation of parasitic polarization effects of the MW at the ion position. 

The letter is organized as follows: After sketching the experimental setup, we describe the spectroscopic sequence for obtaining optical and microwave spectra. Central is the detailed comparison to the theoretical model where we take the dressing and the quadrupol shift due to the Paul potential into account. This leads to a conclusion and outlook how to improve in future the application of MW dressing fields on ion crystals.

\section{Experimental setup}
A single ion is trapped in a linear segmented Paul trap, consisting of 11 gold-coated electrode pairs for ion trapping by DC voltages and one electrode pair for the RF drive, symmetrically arranged around the trap axis. The distance from the RF electrode to the center of the trap structure is 750~$\mu$m. The segments on the DC electrodes are separated by insulating gaps of 30 $\mu m$, while the width of the segments itself varies from 500 $\mu m$ to 1000 $\mu m$. The DC voltages are at the scale of $\pm$ 10~V and the amplitude of the RF voltage is about 500~V$_{pp}$. This RF electric field for generating the dynamical confinement is $\omega_{RF}=2\pi\times 14.1~\text{MHz}$. The resulting secular trap frequencies in three directions are $\{\omega_x,\omega_y,\omega_z\}=2\pi\times\{1.78,2.10,1.07\}~\text{MHz}$. We use to Doppler cool the $^{40}$Ca$^+$ ion to an average phonon number near 10 by a laser source near 397~nm, slightly red detuned from the transition $4S_{1/2}\rightarrow 4P_{3/2}$. We further cool the ion by the resolved sideband cooling on the quadrupole transition $4S_{1/2}\leftrightarrow 4D_{5/2}$ with a laser near 729~nm laser. Close to the ground state of vibration, any shift or broadening of the Rydberg excitation is suppressed. We employ two laser sources near 213~nm and 287~nm. Additionally, we can apply a microwave near 90~GHz. This is generated as the 7$^{th}$ harmonic of a low phase noise MW generator, emitted by a horn antenna and steered via an off-axis parabola mirror into a viewport of the trap vacuum chamber. We control the polarization by a wire-based polarizer. All laser sources and the microwave source can switched on and off by computer control to drive a spectroscopy sequence. We collect laser-induced fluorescence light by a high NA objective and steer this to an EMCCD camera for detection. 

\section{Energy levels in Calcium and sequence for spectroscopy}

\begin{figure}
\includegraphics[width=0.95\linewidth]{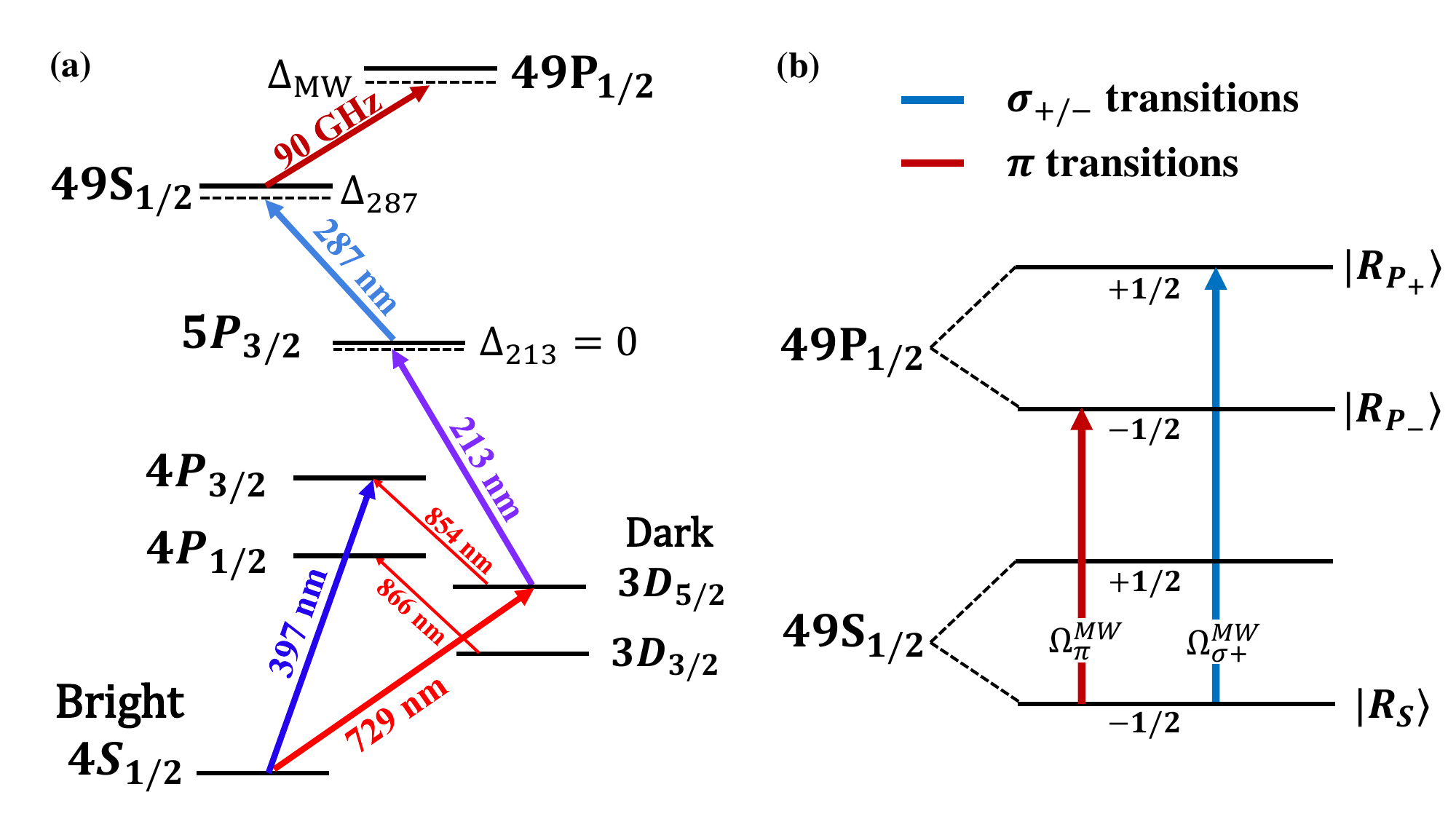}
\caption{\label{levels}Levels and transitions of $^{40}Ca^+$. (a) An ion in the state $|B\rangle=|4S_{1/2}\rangle$ is excited with a laser near 397~nm and detected via laser-induced fluorescence. The laser sources near 866~nm and 854~nm pump the ion from the metastable D levels to $|B\rangle$. The laser near 729~nm is employed to coherently transfer the ion state to the metastable state $|D\rangle=|3D_{5/2},m=-5/2\rangle$. The electromagnetically-induced-transparency (EIT) ladder-configuration is established between $|D\rangle$, the intermediate state $|I\rangle=|5P_{3/2},m=-3/2\rangle$ and the Rydberg S-state $|R_S\rangle=|49S_{1/2},m=-1/2\rangle$ coupled by lasers near 213~nm and 287~nm, respectively. The Rydberg S- and P-states are dressed by the microwave field near 90~GHz.  (b) Zeeman sub-levels of the Rydberg states and microwave coupling: the state $|R_S\rangle$ is excited by the laser sources, and coupled by $\pi$ polarization (red) to $|R_{P_-}\rangle=|49P_{1/2},m=-1/2\rangle$, or by $\sigma+$ polarization (blue) to the $|R_{P_+}\rangle=|49P_{1/2},m=+1/2\rangle$ state.}
\end{figure}

\begin{figure*}[h]
\includegraphics[width=0.95\linewidth]{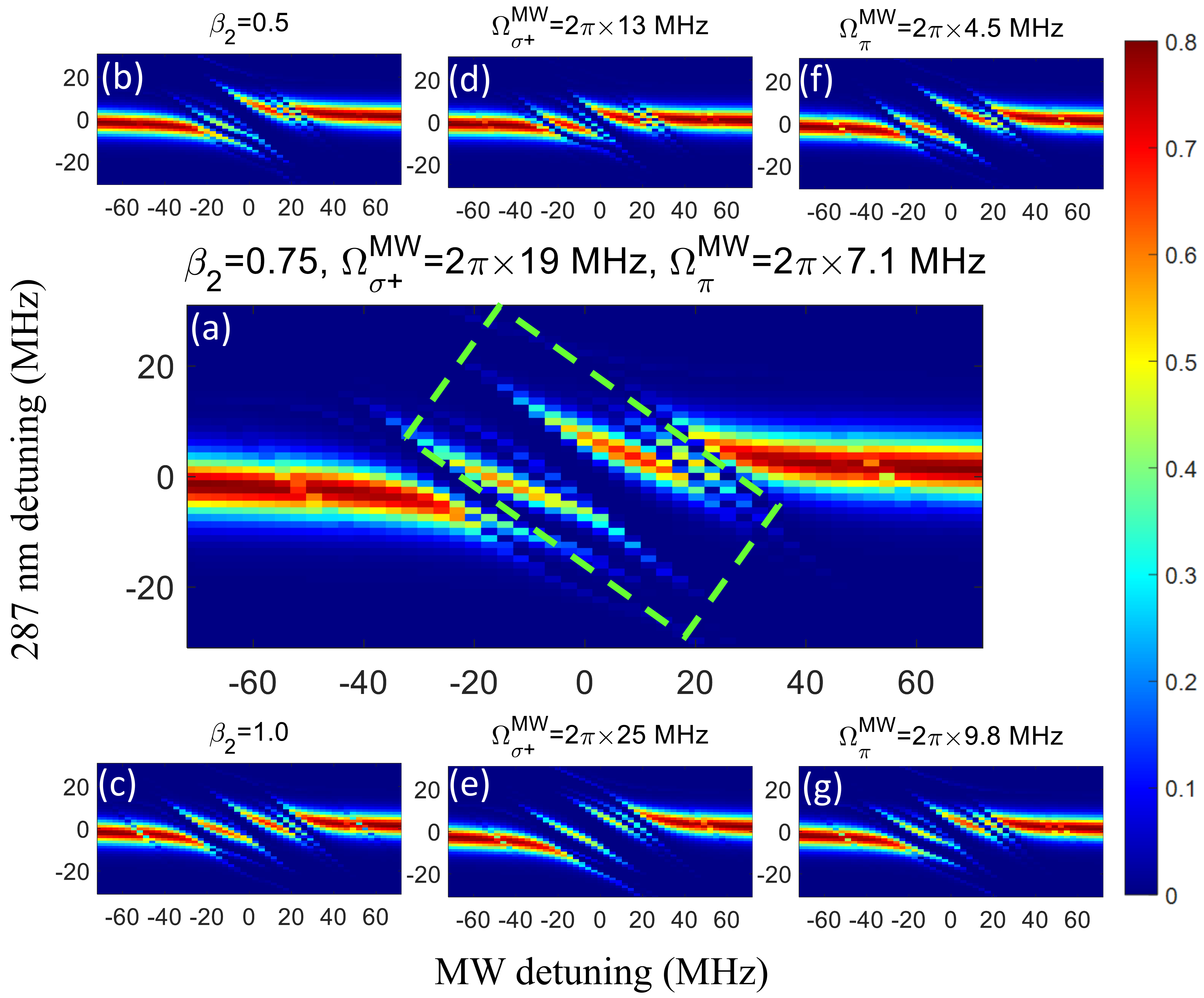}
\caption{\label{theory}Numerical results of the dark state probability after an evolution time of $150~\mu s$ for different choices of the parameters \textbf{(a)} Numerical result with optimized parameters $\beta_2=0.75, \Omega_{\sigma+}^{MW}=2\pi\times19~\text{MHz}, \Omega_{\pi}^{MW}=2\pi\times7.1~\text{MHz}$. The green dashed rectangle in the center is the region of data used for optimizing the parameters. \textbf{(b-c)} Numerical results with $\beta_2$ adjusted while controlling the other two parameters. \textbf{(d-e)} Numerical results with $\Omega_{\sigma+}^{MW}$ adjusted while controlling the other two parameters. \textbf{(f-g)} Numerical results with $\Omega_{\pi}^{MW}$ adjusted while controlling the other two parameters.}
\end{figure*}

The energy levels of $^{40}$Ca$^+$, see Fig.~\ref{levels} (a), allow for detection, when the ion is in the bright state $|B\rangle=|4S_{1/2}\rangle$. The laser source near 397~nm is exciting to the $|4P_{1/2}\rangle$ state and the fluorescence is detected. If the ion is transferred to the long-lived (1.17 s) meta-stable dark state $|D\rangle=|3D_{5/2},m=-5/2\rangle$ by a laser near 729~nm, no resonance fluorescence is emitted. From the metastable state, and using laser sources near 213~nm and 287~nm lasers, we coupled to the Rydberg S state $|R_S\rangle=|49S_{1/2},m=-1/2\rangle$ via an intermediate state $|I\rangle=|5P_{3/2},m=-3/2\rangle$. The laser near 213~nm s tuned on resonance to the state $|I\rangle$. Since this state has a short lifetime of 34.8~ns, this will rapidly pump the ion to $|B\rangle$. Subsequently, the ion will be measured as bright, as it emits fluorescence when excitetd by the laser near 397~nm. However, if a strong second laser near 287~nm is tuned to be on resonant with the transition from $|I\rangle$ to $|R_S\rangle$, the Autler-Townes (AT) effect~\cite{Cohen-Tannoudji1996,hao2018transition,Abi-Salloum2010} will shift the first excitation out of resonance, thus reducing the optical pumping rate. The ion electronic state will remain in $|D\rangle$. Moreover, if the $|R_S\rangle$ state is coupled to $49P_{1/2}$ by a $\sim$90 GHz microwave, this coupling is inducing an AT effect for the transition $|I\rangle\leftrightarrow|R_S\rangle$ to make it off-resonant. Therefore, this affects the Autler-Townes effect for the transition $|D\rangle\leftrightarrow|I\rangle$, and is thus reducing the optical pumping rate to the electronic ground state $|B\rangle$. 

Taking advantage of the high fidelity state detection of a single ion either in $|B\rangle$, with many fluorescence photons detected, or in $|D\rangle$, no photon detection, we can realize combined optical and microwave spectroscopy. The experiment sequence is as follows: Doppler cooling for 10~ms, followed by resolved sideband cooling for 2.3~ms, a transfer pulse near 729~nm to the $|D\rangle$ state by a rapid adiabatic passage (RAP) within 0.2~ms, and the Rydberg coupling with lasers near 213~nm and 287~nm laser together with a rectangular microwave pulse for 150~$\mu$s.

The microwave polarization determines how the state $|R_S\rangle$ couples to Zeeman sub-levels of the $49P_{1/2}$, that we denote by $|R_{P_-}\rangle$ and $|R_{P_+}\rangle$, see Fig.~\ref{levels} (b). Using a grid wire polarizer, the microwave polarization is set to be linear. The microwave propagates in the x-z plane and impinges the ion at a $\pi/4$ angle to the trap z-axis. We apply a magnetic field of 560~mT in direction of the microwave propagation to simplify the coupling scheme, as the linear polarized field in decomposed into $\sigma_+$ and $\sigma_-$ polarization only. The state $|R_S\rangle$, which is excited by the optical fields, is coupled by $\sigma_+$ polarization with a strength of $\Omega^{MW}_{\sigma+}$ to $|R_{P_+}\rangle$ (blue arrow), but the $\pi$-polarization should not be contained in the field leading to an $\Omega^{MW}_{\pi}$ close to zero. 

\section{Theory Model of Microwave-dressed Rydberg States}

We initialize the ion elecronic state in $|D\rangle$. Considering the energy levels interacting with the laser- and microwave-fields, the Hamiltonian after rotating wave approximation can be written as
\begin{equation}
\begin{aligned}
H=H_{E}+H_R
\end{aligned}
\end{equation}
where $H_{E}$ describe the interaction within the ladder configuration including the dark state $|D\rangle$, the intermediate state $|I\rangle$ and the Rydberg S state $|R_S\rangle$.
\begin{equation}
\label{static Hamiltonian}
\begin{aligned}
H_{E}=&\frac{\hbar}{2}(\Omega_{213}|D\rangle\langle I|+\Omega_{287}|I\rangle\langle R_S|)+h.c.\\
-&\hbar\Delta_{287}|R_S\rangle\langle R_S|
\end{aligned}
\end{equation}
$\Omega_{213}$ and $\Omega_{287}$ denote the Rabi frequencies of the laser drive near 213~nm and 287~nm, see Fig. \ref{levels}. $\Delta_{287}$ is the detuning of the laser near 287~nm with respect to the transition $|I\rangle\leftrightarrow|R_S\rangle$, while we set the detuning of the laser near 213~nm with respect to the transition $|D\rangle\leftrightarrow|I\rangle$  to zero. The microwave mediated interaction between $|R_S\rangle$ and Rydberg P levels $|R_{P_{+,-}}\rangle=|49P_{1/2},m=\pm 1/2\rangle$ writes
\begin{equation}
\begin{aligned}
H_R=&\hbar(H_{R_{P_+}}+H_{R_{P_-}})\\
=&\hbar\left(\frac{\Omega_{\sigma+}^{MW}}{2}|R_{P_+}\rangle\langle R_S|+h.c.-\Delta_+|R_{P_+}\rangle\langle R_{P_+}|\right)\\
+&\hbar\left(\frac{\Omega_{\pi}^{MW}}{2}|R_{P_-}\rangle\langle R_S|+h.c.-\Delta_-|R_{P_-}\rangle\langle R_{P_-}|\right)
\end{aligned}
\end{equation}

Since we have carefully adjusted the microwave polarization and propagation direction with respect to the magnetic field, the microwave should only couple the transition $|R_S\rangle\leftrightarrow|R_{P_+}\rangle$. Solving the master equations disregarding $|R_{P_-}\rangle$, we get the trivial avoided crossing configuration of the dressed state between $|R_S\rangle$ and $|R_{P_+}\rangle$ as shown in Fig.~\ref{theory}(a).

However, the experiment result features a more complex pattern. We find out that this originates from the electric quadrupole modulation, where the Rydberg state of the ion experiences the RF-oscillating electric field of the Paul trap at $\omega_{RF}=2\pi\times 14.1$~MHz. The spatial distribution of the trap drive is in the form of electric quadrupole. Due to the quadrupole polarizability of $49P_{1/2}$, the ion energy level is modulated with amplitude $E_1$ and $E_2$ at frequency $\omega_{RF}$ and $2\omega_{RF}$~\cite{Hig2021,Feldker2015PRL}. Since $E_2\gg E_1$, the extra oscillating term can be simplified to
\begin{equation}
\label{oscillating Hamiltonian}
\begin{aligned}
H_M(t)=E_2\cos{(2\omega_{RF}t)}\left(|R_{P_+}\rangle\langle R_{P_+}|+|R_{P_-}\rangle\langle R_{P_-}|\right)
\end{aligned}
\end{equation}
We can solve the master equation by Trotter formulas\cite{Trotter}, which means decomposing the evolution operator into the product of the operators within small time slices as
\begin{equation}
\begin{aligned}
U_{all}&=\hat{P}\exp{\left(-\frac{i}{\hbar}\int_0^{t}(H+H_{M}(\tau))d\tau\right)}\\
&\approx\prod_j\exp{\left(-\frac{i}{\hbar}(H+H_{M}(\tau_j))\Delta\tau\right)}
\end{aligned}
\end{equation}
where $\hat{P}$ is the time order operator, $\Delta\tau$ the interval of the time slices and $\tau_j$ the time of the j-th slice. In another perspective, we can go into the interaction picture of $H_M$. The corresponding unitary operator can be expanded as
\begin{equation}
\begin{aligned}
U_M=&e^{i\int_0^t H_M\tau d\tau/\hbar}\\
=&\sum_k\mathcal{J}_k(\beta_2)e^{i2k\omega_{RF}t}\left(|R_{P_+}\rangle\langle R_{P_+}|+|R_{P_-}\rangle\langle R_{P_-}|\right)
\end{aligned}
\end{equation}

where $\beta_2=\frac{E_2}{2\hbar\omega_{RF}}$ denotes the modulation depth and $\mathcal{J}_k$ is the k-th Bessel function of the first kind. Finally, the microwave dressing terms in the interaction picture become
\begin{equation}
\begin{aligned}\label{interaction_picture}
U_MH_{R}U_M^{\dagger}=&-\hbar \left(\Delta_+|R_{P_+}\rangle\langle R_{P_+}|+\Delta_-|R_{P_-}\rangle\langle R_{P_-}|\right)\\
+&\hbar\mathcal{J}_k(\beta_2)\frac{\Omega_{\sigma+}^{MW}e^{i2k\omega_{RF}t}}{2}|R_{P_+}\rangle\langle R_S|+h.c.\\
+&\hbar\mathcal{J}_k(\beta_2)\frac{\Omega_{\pi}^{MW}e^{i2k\omega_{RF}t}}{2}|R_{P_-}\rangle\langle R_S|+h.c.
\end{aligned}
\end{equation}
As a result, we see that the modulation of the Rydberg P-state energy can be treated as if the microwave would carry multiple frequency components with $2\omega_{RF}$ intervals. For a typical value in our experiment with $\beta_2\approx 0.8$, the contributions are $|\mathcal{J}_2(\beta_2)|\approx0.031$ and $|\mathcal{J}_3(\beta_2)|\approx0.0026$. Therefore, the amplitudes of the high order frequency modulation components are negligible.

\begin{figure}
\includegraphics[width=0.95\linewidth]{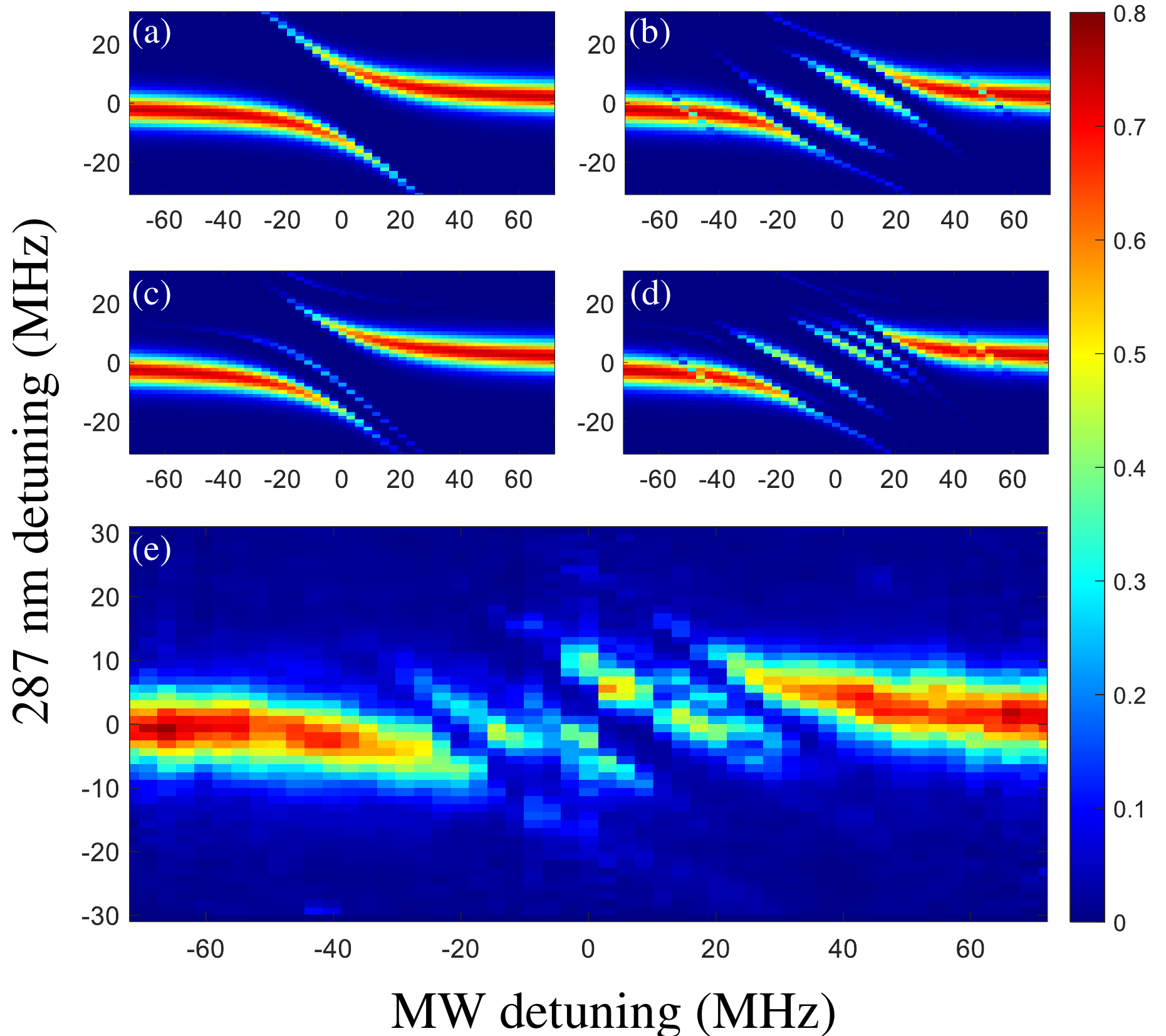}
\caption{\label{exp}\textbf{(a)} The dressed states between $|R_S\rangle$ and $|R_{P_+}\rangle$ coupled with pure $\sigma_+$ polarization, where we disregard quadrupole modulation. \textbf{(b)} The dressed states between $|R_S\rangle$, $|R_{P_-}\rangle$ and $|R_{P_+}\rangle$, where a weak parasitic $\pi$ polarization of the microwave field is included. \textbf{(c)} The dressed states, now with considering the quadrupole modulation, for pure $\sigma_+$ polarization. \textbf{(d)} Same as \textbf{(c)} but with a microwave-field including both $\sigma_+$ polarization, and a weak residual $\pi$ polarization. The parameters used in \textbf{(d)} are: $\beta_2=0.8$, $\Omega_{\sigma+}^{MW}=2\pi\times 25~\text{MHz}$, $\Omega_{\pi}^{MW}=2\pi\times 8.9~\text{MHz}$, $\Omega_{213}=2\pi\times 0.23~\text{MHz}$, $\Omega_{287}=2\pi\times 12~\text{MHz}$.} \textbf{(e)} Experimental results of the dark state probability after $150~\mu s$ of interaction.
\end{figure}

\section{Experimental Results and Comparison to the Theory}
The experimental results are shown in Fig.~\ref{exp}(b). The patterns on the left and right side are similar to the trivial avoided crossing configuration since they correspond to the situation that the microwave is far detuned and the microwave AT effect is too weak to disturb the 287 nm AT effect. Therefore, the 213 nm pumping effect is strongly prohibited by the AT effect near 287~nm. As a result, this is preserving a large probability of the ion to be detected in the dark state $|D\rangle$.

For the situation where the microwave frequency is close to resonance, corresponding to the center region of Fig.~\ref{exp}(b), we find two wide, diagonal strips at the center. They can be explained by the quadrupole modulation $H_M(t)$. We know from Eq.~(\ref{interaction_picture}) that the quadrupole modulation is generating  multiple frequency components with $2\omega_{RF}$ intervals. Under certain parameters, the AC Stark shift produced by the different frequency components of microwaves cancels each other out. Therefore, the disturbance for the AT effect near 287~nm is weak, as well the pumping effect for the laser near 213~nm, such that the ion is remaining with a considerable probability in the $|D\rangle$ state. This argument is confirmed by the numerical simulation results, see Fig.~\ref{theory}(c), and compare to Fig.~\ref{theory}(a).

Besides the wide stripes, we observe in the experimental data two additional narrow diagonal lines. They can be explained by the excitation to the $|R_{P_-}\rangle$ by the parasitic microwave $\pi$-polarization. The energy of $|R_{P_-}\rangle$ and $|R_{P_+}\rangle$ are split due to magnetic field and electric field. The coupling to $|R_{P_-}\rangle$ also apply the AT effect to the $|R_S\rangle$, inducing two narrow lines. It is also confirmed by our numerical results as shown in Fig.~\ref{theory}(d), comparing with Fig. \ref{theory}(c). The parasitic residual microwave $\pi$ polarization may be caused by non-parallel near-field component of the microwave, MW-interferences or reflections because of the metallic trap electrodes. From a comparison to the experimental data, we can determine the fraction of parasitic microwave $\pi$-polarization. With parameters $\beta_2=0.8$, $\Omega_{\sigma+}^{MW}=2\pi\times 25~\text{MHz}$, $\Omega_{\pi}^{MW}=2\pi\times 8.9~\text{MHz}$, $\Omega_{213}=2\pi\times 0.23~\text{MHz}$, $\Omega_{287}=2\pi\times 12~\text{MHz}$, see Fig.~\ref{exp}(d). $\Omega_{213}$ and $\Omega_{287}$ are derived by fitting the corresponding laser spectroscopy data without microwave dressing. 

Tuning the other three parameters $\beta_2$, $\Omega_{\sigma+}^{MW}$ and $\Omega_{\pi}^{MW}$ to make the numerical results consistent with the experiment, all parameters are determined. E.g. the value of $\beta_2$ is determined by the strength of the wide diagonal stripes in the center, the value of $\Omega_{\sigma}^{MW}$ affects how the two main branches on the left and right side bend and the value of $\Omega_{\pi}^{MW}$ determines the strength of the narrow line beside the wide stripes. 

To determine how accurate the parameters are, we select the central region in the numerical results (green dashed rectangle shown in Fig.\ref{theory}(a)) and compare with the experiment outcome. We bin the data in the left up direction  direction, i.e. the direction parallel to the long side of the rectangle to increase the signal-to-noise ratio. This results in an averaged envelope which reveals the center structure including the wide and narrow spectral features. The envelope from numerical and experimental results are denoted as $V_{num}$ and $V_{exp}$, respectively. We use the difference between these two envelopes $D=\sum((V_{num}-V_{exp})^2)$ to quantify the fit of theory with the experimental data, and the minimal value $min(D)=0.065$ 
is reached with the optimized parameters, see Fig.\ref{theory}(a). The $1-\sigma$ error in the experiment corresponds to the uncertainty of $D$ as $\sigma(D)=0.007$. Therefore, we can assign an uncertainty of the parameters as $\beta_2=0.75_{-0.11}^{+0.08}$, $\Omega_{\sigma+}^{MW}=2\pi\times 19_{-1.5}^{+2.0}~\text{MHz}$, $\Omega_{\pi}^{MW}=2\pi\times 7.1_{-1.3}^{+1.2}~\text{MHz}$.


\section{Conclusion and Outlook}
We obtained MW-mixing of Rydberg states in an experiment with a single trapped cold ion and obtain Autler-Townes spectra. This  method provides spectroscopic data, but does not suffer from black-body photon induced ion losses, even in our room-temperature experiment. We model the AT spectra theoretically, including the optical multi-photon Rydberg excitation, the microwave dressing, and the RF quadrupole modulation by the dynamical potential of the Paul trap. We are able to fit the experimental results with our model, and we deduce from this comparison all experimental parameter values accurately. The entire data taking runtime was five days, and therefore we conclude that the overall stability of the setup is good, but notably also the long-term stability of the MW delivery to ions is sufficiently high to maintain the parameters stable. From our data we found that even thought a MW-beam  with proper polarization is injected, the vicinity of trap electrodes seemingly affect the microwave polarization purity. Therefore we conjecture that in future, microwave antenna which are integrated in a trap chip design could be of advantage. Moreover, we plan to suppress the quadrupole modulation by increasing the radial confinement to achieve a modulation index of $\beta_2=3.8$, which marks a zero of the corresponding Bessel function $\mathcal{J}_{\pm1}(\beta_2)\approx 0$. For such parameter, first-order frequency components disappear at the cost of moderate second-order components\cite{Hig2021,simeonov2019}. On a longer run, the use of a Penning trap for Rydberg excitation of ion crystals would mitigate the effect of Rydberg-quadrupole modulation, because in such a trap only static electric and static magnetic fields are used for the confinement. 

\acknowledgments
We thank Dr. Jonas Vogel for discussions. We acknowledge financial support by the German Science Foundation (DFG) within the SPP 1929 Giant interactions in Rydberg Systems (GiRyd) and the ERC Synergy funding within the project "Open 2D Quantum Simulator". 

\bibliographystyle{unsrt}
\bibliography{main}

\begin{thebibliography}{10}

\bibitem{Klepp1995}
Michael Courtney, Neal Spellmeyer, Hong Jiao, and Daniel Kleppner.
\newblock Classical, semiclassical, and quantum dynamics in the lithium {S}tark
  system.
\newblock {\em Phys. Rev. A}, 51:3604--3620, May 1995.

\bibitem{GALLAGHER2008161}
Thomas~F. Gallagher and Pierre Pillet.
\newblock Dipole–dipole interactions of {R}ydberg atoms.
\newblock In {\em Advances in Atomic, Molecular, and Optical Physics},
  volume~56 of {\em Advances In Atomic, Molecular, and Optical Physics}, pages
  161--218. Academic Press, 2008.

\bibitem{Zha2020}
Chi Zhang, Fabian Pokorny, Li~Weibin, Gerard Higgins, Andreas Pöschl, Igor
  Lesanowsky, and Markus Hennrich.
\newblock Submicrosecond entangling gate between trapped ions via {R}ydberg
  interaction.
\newblock {\em Nature}, 580:345, 2020.

\bibitem{Li2013}
Weibin Li and Igor Lesanowsky.
\newblock Entangling quantum gate in trapped ions via {R}ydberg blockade.
\newblock {\em Appl. Phys. B}, 114:37, 2014.

\bibitem{Cohen-Tannoudji1996}
Claude~N. Cohen-Tannoudji.
\newblock {\em The Autler-Townes Effect Revisited}, pages 109--123.
\newblock Springer New York, New York, NY, 1996.

\bibitem{hao2018transition}
Liping Hao, Yuechun Jiao, Yongmei Xue, Xiaoxuan Han, Suying Bai, Jianming Zhao,
  and Georg Raithel.
\newblock Transition from electromagnetically induced transparency to
  {A}utler--{T}ownes splitting in cold cesium atoms.
\newblock {\em New Journal of Physics}, 20(7):073024, 2018.

\bibitem{Abi-Salloum2010}
Tony~Y. Abi-Salloum.
\newblock Electromagnetically induced transparency and {A}utler-{T}ownes
  splitting: Two similar but distinct phenomena in two categories of
  three-level atomic systems.
\newblock {\em Phys. Rev. A}, 81:053836, May 2010.

\bibitem{Hig2021}
Gerard Higgins, Chi Zhang, Fabian Pokorny, Harry Parke, Erik Jansson, Shalina
  Salim, and Markus Hennrich.
\newblock Observation of second- and higher-order electric quadrupole
  interactions with an atomic ion.
\newblock {\em Phys. Rev. Res.}, 3:L032032, Aug 2021.

\bibitem{Feldker2015PRL}
T.~Feldker, P.~Bachor, M.~Stappel, D.~Kolbe, R.~Gerritsma, J.~Walz, and
  F.~Schmidt-Kaler.
\newblock Rydberg excitation of a single trapped ion.
\newblock {\em Phys. Rev. Lett.}, 115:173001, Oct 2015.

\bibitem{Trotter}
H.~F. Trotter.
\newblock On the product of semi-groups of operators.
\newblock {\em Proceedings of the American Mathematical Society},
  10(4):545--551, 1959.

\bibitem{simeonov2019}
Lachezar~S Simeonov, Nikolay~V Vitanov, and Peter~A Ivanov.
\newblock Compensation of the trap-induced quadrupole interaction in trapped
  {R}ydberg ions.
\newblock {\em Scientific Reports}, 9(1):7340, 2019.

\end{thebibliography}

\end{document}